\pdfoutput=1
\documentclass[sigconf]{acmart}
\usepackage{mathrsfs}
\usepackage{chngcntr}
\usepackage{titlesec}
\usepackage[english]{babel}  
\usepackage{amssymb}
\usepackage{arydshln}  
\usepackage{xparse}
\usepackage[shortlabels]{enumitem}
\usepackage{boxedminipage}
\usepackage[capitalise]{cleveref}
\crefname{paragraph}{Section}{Sections}
\Crefname{paragraph}{Section}{Sections}

\setcopyright{rightsretained}
\copyrightyear{2019}
\acmYear{2019}
\acmConference[ISSAC '19]{International Symposium on Symbolic and AlgebraicComputation}{July 15--18, 2019}{Beijing, China}
\acmBooktitle{International Symposium on Symbolic and Algebraic Computation(ISSAC '19), July 15--18, 2019, Beijing, China}
\acmPrice{15.00}
\acmDOI{10.1145/3326229.3326272}
\acmISBN{978-1-4503-6084-5/19/07}
\makeatletter
\def\renewtheorem#1{%
  \expandafter\let\csname#1\endcsname\relax
  \expandafter\let\csname c@#1\endcsname\relax
  \gdef\renewtheorem@envname{#1}
\renewtheorem@secpar}
\def\renewtheorem@secpar{\@ifnextchar[{\renewtheorem@numberedlike}{\renewtheorem@nonumberedlike}}
\def\renewtheorem@numberedlike[#1]#2{\newtheorem{\renewtheorem@envname}[#1]{#2}}
\def\renewtheorem@nonumberedlike#1{  
  \def\renewtheorem@caption{#1}
  \edef\renewtheorem@nowithin{\noexpand\newtheorem{\renewtheorem@envname}{\renewtheorem@caption}}
  \renewtheorem@thirdpar
}
\def\renewtheorem@thirdpar{\@ifnextchar[{\renewtheorem@within}{\renewtheorem@nowithin}}
\def\renewtheorem@within[#1]{\renewtheorem@nowithin[#1]}

\newtheoremstyle{framedthmenv}%
{0cm}
{0cm}
{\@acmdefinitionbodyfont}
{0cm}
{\@acmdefinitionheadfont}
{:}
{.5em}
{\thmname{#1}\thmnumber{ #2}\thmnote{ {\@acmdefinitionnotefont(#3)}}}
\makeatother

\theoremstyle{acmplain}
\renewtheorem{theorem}{Theorem}[section]
\renewtheorem{proposition}[theorem]{Proposition}
\renewtheorem{lemma}[theorem]{Lemma}
\renewtheorem{corollary}[theorem]{Corollary}
\theoremstyle{acmdefinition}
\renewtheorem{definition}[theorem]{Definition}

\theoremstyle{framedthmenv}
\crefname{problem}{Problem}{Problems}
\Crefname{problem}{Problem}{Problems}
\newtheorem{algorithm}{Algorithm} 
\crefname{algorithm}{Algorithm}{Algorithms}
\Crefname{algorithm}{Algorithm}{Algorithms}

\crefname{remark}{Remark}{Remarks}
\Crefname{remark}{Remark}{Remarks}
\theoremstyle{acmplain}

\newcommand{\bigO}[1]{O(#1)} 
\newcommand{\softO}[1]{\mathchoice{\tilde{O}\left(#1\right)}{O\tilde{~}(#1)}{O\tilde{~}(#1)}{O\tilde{~}(#1)}} 
\newcommand{\expmm}{\omega} 
\newcommand{\algoname}[1]{{\normalfont\textsc{#1}}} 
\newcommand{\algoword}[1]{\emph{\textsf{#1}}} 
\newcommand{\assign}{\leftarrow} 
\newcommand{\ZZ}{\mathbb{Z}} 
\newcommand{\ZZp}{\mathbb{Z}_{> 0}} 
\newcommand{\field}{\mathbb{K}} 
\newcommand{\order}{d} 

\def\M{\mathsf{M}}
\def\MM{\mathsf{MM}}

\def\N {\ensuremath{\mathsf{N}}}
\newcommand{\var}{x} 
\newcommand{\polRing}{\field[\var]} 
\newcommand{\matRing}[2]{\field^{#1 \times #2}} 
\newcommand{\pmatRing}[2]{\polRing^{#1 \times #2}} 
\newcommand{\row}[1]{\boldsymbol{#1}} 
\newcommand{\col}[1]{\boldsymbol{#1}} 
\newcommand{\mat}[1]{\mathbf{\MakeUppercase{#1}}} 
\newcommand{\trsp}[1]{#1^\mathsf{T}} 
\newcommand{\ident}[1]{\mat{I}_{#1}} 

\newcommand{\tuple}[1]{\boldsymbol{#1}} 
\newcommand{\shifts}{\tuple{s}} 
\newcommand{\shiftss}{\tuple{t}} 
\newcommand{\rdeg}[2]{\mathrm{rdeg}_{#1}(#2)} 

\newlist{algosteps}{enumerate}{3}
\crefname{algostepsi}{Step}{Steps}
\crefname{algostepsi}{Step}{Steps}
\Crefname{algostepsii}{Step}{Steps}
\Crefname{algostepsii}{Step}{Steps}
\Crefname{algostepsiii}{Step}{Steps}
\Crefname{algostepsiii}{Step}{Steps}
\newenvironment{algobox}{
  \newcommand{\algoInfo}[1]{
    \begin{algorithm}
      \emph{\algoname{##1}}
    }
    
  \newcommand{\dataInfo}[2]{
  \algoword{##1:} ##2 }
  \newcommand{\algoSteps}[1]{
    \setlist[algosteps,1]{leftmargin=0.5cm}
    \setlist[algosteps,2]{leftmargin=0.4cm}
    \setlist[algosteps,3]{leftmargin=0.4cm}
    \begin{algosteps}[label=\textbf{\arabic*.},ref=\arabic*]
      ##1
    \end{algosteps}
  }
  \begin{figure}[ht]
    \centering
    \addtolength\fboxsep{0.03cm}
    \begin{boxedminipage}{0.99\columnwidth}
    }
    {
    \end{algorithm}
  \end{boxedminipage}
\end{figure}
}

\def\MM{\mathsf{MM}}
\def\M{\mathsf{M}}
\def\F{\mathbb{F}}
\def\N{\mathbb{N}}
\def\K{\mathbb{K}}

\def\Z{\mathbb{Z}}
\def\Q{\mathbb{Q}}

\counterwithin{paragraph}{section} 
\counterwithin{paragraph}{subsection} 

\titleformat{\paragraph}[runin]{\normalfont\normalsize\bfseries}{\theparagraph.}{1em}{}
\titlespacing*{\paragraph}{0em}{0.5ex}{0.5em}

\renewcommand{\theparagraph}{%
  \ifnum\value{subsection}=0
    \thesection
  \else
    \thesubsection
  \fi
.\arabic{paragraph}}

\begin{document}
\title{Implementations of efficient univariate polynomial matrix algorithms and
application to bivariate resultants}

\author{Seung Gyu Hyun}
\affiliation{%
  \institution{{\normalsize University of Waterloo}}
  \city{Waterloo, ON} 
  \state{Canada} 
}

\author{Vincent Neiger}
\affiliation{%
  \institution{{\normalsize Univ. Limoges, CNRS, XLIM, UMR\,7252}}
  \city{F-87000 Limoges} 
  \state{France} 
}

\author{\'Eric Schost}
\affiliation{%
  \institution{{\normalsize University of Waterloo}}
  \city{Waterloo, ON} 
  \state{Canada} 
}

\begin{abstract}
  Complexity bounds for many problems on matrices with univariate polynomial
  entries have been improved in the last few years. Still, for most related
  algorithms, efficient implementations are not available, which leaves open
  the question of the practical impact of these algorithms, e.g. on
  applications such as decoding some error-correcting codes and solving
  polynomial systems or structured linear systems. In this paper, we discuss
  implementation aspects for most fundamental operations: multiplication,
  truncated inversion, approximants, interpolants, kernels, linear system
  solving, determinant, and basis reduction. We focus on prime fields with a
  word-size modulus, relying on Shoup's C++ library NTL. Combining these new
  tools to implement variants of Villard's algorithm for the resultant of
  generic bivariate polynomials (ISSAC 2018), we get better performance than
  the state of the art for large parameters.
\end{abstract}

%


\maketitle


\vspace{-0.15cm}
\section{Introduction}
\label{sec:intro}

Hereafter, \(\field\) is a field and \(\polRing\) is the algebra of univariate
polynomials over \(\field\). Recent years have witnessed a host of activity on
fast algorithms for polynomial matrices and their applications:
\begin{itemize}[nosep]
  \item Minimal approximant bases \cite{GiJeVi03,ZhoLab12} were used
    to compute kernel bases \cite{ZhLaSt12}, giving the first
    efficient deterministic algorithm for linear system solving
    over~\(\polRing\).
  \item Basis reduction \cite{GiJeVi03,GuSaStVa12} played a key role
    in accelerating the decoding of one-point Hermitian codes
    \cite{NieBee15} and in designing deterministic determinant and
    Hermite form algorithms~\cite{LaNeZh17}.
  \item Progress on minimal interpolant bases \cite{JeNeScVi17,JeNeScVi16} led
    to the best known complexity bound for list-decoding Reed-Solomon codes and
    folded Reed-Solomon codes \cite[Sec.\,2.4~to~2.7]{JeNeScVi17}.
  \item Coppersmith's block Wiedemann algorithm and its
    extensions~\cite{Coppersmith94,Kaltofen95,Villard97} were used in
    a variety of contexts, from integer factorization~\cite{cadonfs} 
    to polynomial system solving~\cite{Villard18,HyNeRaSc17}.
\end{itemize}
\noindent At the core of these improvements, 
one also finds techniques such as high-order lifting \cite{Storjohann03} and
partial linearization \cite{Storjohann06},\cite[Sec.\,6]{GuSaStVa12}.

For many of these operations, no implementation of the latest algorithms is
available and no experimental evidence has been given regarding their practical
behavior. Our goal is to partly remedy this issue, by providing and discussing
implementations for a core of fundamental algorithms such as multiplication,
approximant and interpolant bases, etc., upon which one may implement higher
level algorithms. As an illustration, we describe the performance of slightly
modified versions of Villard's recent breakthroughs on bivariate resultant and
characteristic polynomial computation~\cite{Villard18}.

Our implementation is based on Shoup's Number Theory Library (NTL) \cite{NTL},
and is dedicated to polynomial matrix arithmetic over $\field = \F_p$ for a
word-size prime $p$.  Particular attention was paid to performance issues, so
that our library compares favorably with previous work for those operations
where comparisons were possible. Our code is available at
\url{https://github.com/vneiger/pml}.

\paragraph*{Overview.}

Polynomial matrix algorithms rely on efficient arithmetic in $\K[x]$ and for
matrices over $\K$; in \cref{sec:arith}, we review some related algorithms and
their NTL implementations. Then, we describe our implementation of a key
building block: multiplication. 

\cref{sec:appint} presents the next major part of our work, concerning
algorithms for \emph{approximant bases}
\cite{BecLab94,GiJeVi03,ZhoLab12,JeNeVi2018} and \emph{interpolant bases}
\cite{BarBul92,BeLa2000,JeNeScVi16,JeNeScVi17}. We focus on a version of
interpolants which is less general than in these references but allows for a
more efficient algorithm. In particular, we show that with this version, both
interpolant and approximant bases can be used interchangeably in several
contexts, with interpolants sometimes achieving better performance than
approximants.  In \cref{sec:higher}, we discuss algorithms for minimal kernel
bases, linear system solving, determinant, and basis reduction. Finally, using
these tools, we study the practical behavior of the bivariate resultant
algorithm of \cite{Villard18} (\cref{sec:Villard}).

Below, cost bounds are given in an algebraic complexity model, counting all operations
in the base field at unit cost. While standard, this point of view fails to
describe parts of the implementation (CRT-based algorithms, such as the
3-primes FFT, cannot be described in such a manner), but we believe that this
is a minor issue.

\paragraph*{Implementation choices.}


NTL is a C++ library for polynomial and matrix arithmetic over rings such as
$\Z$, $\Z/n\Z$, etc., and is often seen as a reference point for fast
implementations in such contexts. Other libraries for these operations include
for example FLINT \cite{flint} as well as FFLAS-FFPACK and
LinBox~\cite{fflas-ffpack,LinBox}. Currently, our implementation relies solely
on NTL; this choice was based on comparisons of performance for the
functionalities we need. \\
\indent In our implementation, the base field is a prime finite field $\F_p$; we rely
on NTL's \texttt{lzz\_p} class. At the time of writing, on standard x86\_64
platforms, NTL~v11.3.1 uses \texttt{unsigned long}'s as its primary data type
for \texttt{lzz\_p}, supporting moduli up to 60 bits long. 

For such fields, one can directly compare running times and cost bounds, since
in the literature most polynomial matrix algorithms are analyzed in the
algebraic complexity model. Besides, computations over \(\F_p\) are at the core
of a general approach consisting in solving problems over $\Z$ or $\Q$ by means
of reduction modulo sufficiently many primes, which are chosen so as to satisfy
several, partly conflicting, objectives. We may want them to support Fourier
transforms of high orders. Linear algebra modulo each prime should be fast, so
we may wish them to be small enough to support vectorized matrix arithmetic
with SIMD instructions. On the other hand, using larger primes allows one to
use fewer of them, and reduces the likelihood of unlucky choices in randomized
algorithms.

As a result, while all NTL {\tt lzz\_p} moduli are supported, our
implementation puts an emphasis on three families: small FFT primes that
support AVX-based matrix multiplication (such primes have at most 23 bits);
arbitrary size FFT primes (at most 60 bits); arbitrary moduli (at most 60
bits). Very small fields such as $\F_2$ or $\F_3$ are supported, but we did not
make any specific optimization for them.

\paragraph*{Experiments.}

All runtimes below are in seconds and were measured on an Intel Core i7-4790
CPU with 32GB RAM, using the version 11.3.1 of NTL. Unless specified otherwise,
timings are obtained modulo a random 60 bit prime.  Runtimes were measured on a
single thread; currently, most parts of our code do not explicitly exploit
multi-threading. Tables below only show a few selected timings, with the best
time(s) in bold; for more timings, see
\url{https://github.com/vneiger/pml/tree/master/benchmarks}.


\section{Basic polynomial and matrix arithmetic}
\label{sec:arith}

We review basic algorithms for polynomials and matrices, and related
complexity results that hold over an abstract field $\K$, and we describe how
we implemented these operations. Hereafter, for $d \ge 0$, $\K[x]_d$ is the
set of elements of $\K[x]$ of degree less than $d$.


\paragraph{Polynomial multiplication.}

Multiplication in \(\polRing\) and Fast Fourier Transform (FFT) are
cornerstones of most algorithms in this paper. Let $\M: \N \to \N$ be a
function such that polynomials of degree at most $d$ in \(\polRing\) can be
multiplied in $\M(d)$ operations in $\K$. If $\K$ supports FFT, we can take
$\M(d) \in O(d \log(d))$, and otherwise, $\M(d) \in O(d \log(d) \log\log(d))$
\cite[Chapter~8]{vzGathen13}; as in this reference, we assume that $d\mapsto
\M(d)/d$ is increasing. A useful variant of multiplication is the \emph{middle
product} \cite{HaQuZi04,BoLeSc03}: for integers $c$ and $d$, and $F$ in
$\K[x]_c$ and $G$ in $\K[x]_{c+d}$, $\algoname{MiddleProduct}(F,G,c,d)$ returns
the slice of the product $FG$ with coefficients of degrees $c,\dots,c+d-1$; a
common case is with $c = d$. The direct approach computes the whole product and
extracts the slice.
Yet, the \emph{transposition principle}~\cite{KaKiBs88} yields a more efficient
approach, saving a constant factor (roughly a factor $2$ when $c=d$, if FFT
multiplication is used).


Polynomial matrix algorithms frequently use fast evaluation and interpolation
at multiple points. In general, subproduct tree techniques
\cite[Chapter~10]{vzGathen13} allow one to do evaluation and interpolation of
polynomials in $\K[x]_d$ at $d$ points in $O(\M(d)\log(d))$ operations. For
special sets of points, one can do better: if we know $\alpha$ in $\K$ of order
at least $d$, then evaluation and interpolation at the geometric progression
$(1,\alpha,\ldots,\alpha^{d-1})$ can both be done in time $O(\M(d))$
\cite{BosSch05}. 


In NTL, multiplication in \(\F_p[x]\) uses either naive, Karatsuba, or FFT
techniques, depending on \(p\) and on the degree (NTL provides FFT primes with
roots of unity of order $2^{25}$, and supports arbitrary user-chosen FFT
primes).  FFT multiplication uses the TFT algorithm of~\cite{Harvey2009} and
Harvey's improvements on arithmetic mod $p$~\cite{Harvey2014}. For primes $p$
that do not support Fourier transforms, multiplication is done by means of
either 3-primes FFT techniques \cite[Chapter~8]{vzGathen13} or Sch\"onhage and
Strassen's algorithm. We implemented middle products for naive, Karatsuba and
FFT multiplication, closely following \cite{HaQuZi04,BoLeSc03}, as well as
evaluation/interpolation algorithms for general sets of points and for
geometric progressions.



\paragraph{Matrix multiplication.}

Let $\expmm$ be such that $n \times n$ matrices over any ring can be multiplied
by a bilinear algorithm doing $O(n^\expmm)$ ring operations. The naive
algorithm does exactly $n^3$ multiplications. First improvements due to
Winograd and Waksman \cite{Winograd68,Waksman70} reduced the number of
operations to $n^3/2 + O(n^2)$ if $2$ is a unit.  Strassen's and Winograd's
recursive algorithms \cite{Strassen69,Winograd71} have $\expmm = \log_2(7)$;
the best known bound is $\expmm \le 2.373$ \cite{CopWin90,LeGall14}. Note that,
using blocking, rectangular matrices of sizes $(m \times n)$ and $(n \times p)$
can be multiplied in $O(mnp \min(m,n,p)^{\expmm-3})$ ring operations. NTL
implements its own arithmetic for matrices over $\F_p$ and chooses one of
several implementations depending on the bitsize of $p$, the matrix dimensions,
the available processor instructions, etc.



\paragraph{Polynomial matrix multiplication.}

In what follows, we write $\MM(n,d)$ for a function such that two $n \times n$
matrices of degree at most $d$ can be multiplied in $\MM(n,d)$ operations in
$\K$; we make the assumption that $d \mapsto \MM(n,d)/d$ is
increasing for all $n$.

From the definitions above we obtain $\MM(n,d) \in O(n^\expmm \M(d))$, which is
in $\softO{n^\expmm d}$. Using evaluation/interpolation at
$1,\alpha,\ldots,\alpha^{2d}$ or at roots of unity, one obtains the following
bounds on $\MM(n,d)$:
%
\begin{itemize}[nosep]
  \item $\bigO{n^\expmm d + n^2 \M(d)}$ if an element \(\alpha\) in \(\field\) of
    order more than \(2d\) is known \citep[Thm.\,2.4]{BosSch05}.
  \item $\bigO{n^\expmm d + n^2 d\log(d)}$ if $\K$ supports FFT in degree \(2d\).
\end{itemize}
We also mention a polynomial analogue of an integer matrix multiplication
algorithm from \cite{DoGiLeSc18} which uses evaluation/interpolation, done
plainly via multiplication by (inverse) Vandermonde matrices. Then, the
corresponding part of the cost (e.g. $O(n^2 \M(d))$ for geometric progressions)
is replaced by the cost of multiplying matrices over \(\field\) in sizes
roughly $(d \times d)$ by $(d \times n^2)$; this is in $O(n^2 d^{\expmm-1})$ if
$d \le n^2$. For moderate values of $d$, where $\M(d)$ is not in the FFT
regime, this allows us to leverage fast matrix multiplication over \(\field\).

We implemented and compared various algorithms for matrix multiplication over
\(\F_p[x]\). For matrices of degree less than 5, we use dedicated routines
based on Karatsuba's and Montgomery's formulas~\cite{Montgomery05}; for
matrices of small size (up to 10, depending on \(p\)), we use Waksman's
algorithm.  For other inputs, most of our efforts were spent on variants of the
evaluation/interpolation scheme.

For FFT primes, we use evaluation/interpolation at roots of unity. For general
primes, we use either evaluation/interpolation at geometric progressions (if
such points exist in $\F_p$), or our adaptation of the algorithm
of~\cite{DoGiLeSc18}, or 3-primes multiplication (as for polynomials, we lift
the product from $\F_p[x]$ to $\Z[x]$, where it is done modulo up to 3 FFT
primes). No single variant outperformed or underperformed all others for all
sizes and degrees, so thresholds were experimentally determined to switch
between these options, with different values for small (less than 23 bits) and
for large primes.

Middle product versions of these algorithms were implemented, and are used in
approximant basis algorithms (\cref{ssec:appbas}) and Newton iteration
(\cref{ssec:linsolve}). Multiplier classes are available: they do
precomputations on a matrix $\mat{A}$ to accelerate repeated multiplications by
$\mat{A}$; they are used in Dixon's algorithm (\cref{ssec:linsolve}).

The table below shows timings for our multiplication and LinBox' one, for
random $m\times m$ matrices of degree $d$ and two choices of prime \(p\).  The
global comparison showed running times that are either similar or in favor of
our implementation.
\\
{\small
\begin{tabular}{c|c|ccc|ccc}
      &     & \multicolumn{3}{c|}{20 bit FFT prime} & \multicolumn{3}{c}{60 bit prime} \\
  $m$ & $d$ & ours & Linbox & ratio & ours & Linbox & ratio \\
  \hline
  8	& 131072	& \textbf{1.1198}	& 1.5930	& 0.70 & \textbf{3.577} & 13.59 & 0.26 \\
  32	& 4096	& \textbf{0.4283}	& 0.5092	& 0.84 & \textbf{2.000} &5.330	&0.38\\
  128	& 1024	& \textbf{1.7292}	& 2.1126	& 0.82 & \textbf{15.73} & 23.13 & 0.68 \\
  512	& 128	& \textbf{4.3533}	& 4.3837	& 0.99 & \textbf{41.57} & 50.62 & 0.82
\end{tabular}
}



\section{Approximant bases and interpolant bases}
\label{sec:appint}

These bases are matrix generalizations of Pad\'e approximation and play an
important role in many higher-level algorithms. For $\mat{F}$ in
$\pmatRing{m}{n}$ and $M$ non-constant in $\K[x]$, they are bases of the
$\K[x]$-module $\mathcal{A}_{M}(\mat{F})$ of all $\row{p}$ in $\K[x]^{1 \times
m}$ such that $\row{p} \mat{F} = 0 \bmod M$. Specifically, {\em approximant bases}
are for \(M = x^\order\) and {\em interpolant bases} for \(M = \prod_i
(\var-\alpha_i)\) for \(\order\) distinct points $\alpha_1, \ldots,
\alpha_\order$ in $\K$. (Here, we do not consider more general cases from the
literature, for example with several moduli \(M_1,\ldots,M_n\), one for each
column of \(\row{p}\mat{F}\).)

Since $\mathcal{A}_{M}(\mat{F})$ is free of rank $m$, such a basis is
represented row-wise by a nonsingular \(\mat{P}\) in \(\pmatRing{m}{m}\). The
algorithms below return \(\mat{P}\) in \emph{\(\shifts\)-ordered weak Popov
form} (also known as \(\shifts\)-quasi Popov form \cite{BeLaVi99}), for a given
\emph{shift} $\shifts=(s_1,\dots,s_m)$ in \(\ZZ^m\). Shifts allow us to set
degree constraints on the sought basis \(\mat{P}\), and they inherently occur
in a general approach for finding bases of solutions to equations
(approximants, interpolants, kernels, etc.).
Approximant basis algorithms often require \(\mat{P}\) to be in
\emph{\(\shifts\)-reduced form} \cite{BarBul92}; although the
\(\shifts\)-ordered weak Popov form is stronger, obtaining it involves minor
changes in these algorithms, without impact on performance according to our
experiments. Recent literature shows that this stronger form reveals valuable
information for further computations with \(\mat{P}\)
\cite{JeNeScVi16,JeNeVi2018}, in particular for finding
\emph{\(\shifts\)-Popov} bases \cite{BeLaVi99}.

From the shift \(\shifts\), the \emph{\(\shifts\)-degree} of $\row{p} =[p_i]_i
\in \K[x]^{1 \times m}$ is defined as $\rdeg{\shifts}{\row{p}}=\max_{1 \le i
\le m} (\deg(p_i) + s_i)$, which extends to matrices: $\rdeg{\shifts}{\mat{P}}$
is the list of \(\shifts\)-degrees of the rows of \(\mat{P}\). Then, the
\emph{\(\shifts\)-pivot} of \(\row{p}\) is its rightmost entry \(p_i\) such
that \(\rdeg{\shifts}{\row{p}} = \deg(p_i) + s_i\), and a nonsingular matrix
\(\mat{P}\) is in \(\shifts\)-ordered weak Popov form if the \(\shifts\)-pivots
of its rows are located on the diagonal.

To simplify cost bounds below, we make use of the function $\MM'(m,d) =
{\sum}_{i=0}^{\log_2(d)} 2^i \MM(m,d/2^i) \in O(\MM(m,d)\log(d))$.


\paragraph{Approximant bases.}
\label{ssec:appbas}

For $\mat{F}$ in $\K[x]^{m \times n}$ and \(d\) in \(\ZZp\), an
\emph{approximant basis} for $(\mat{F},\order)$ is a nonsingular \(m\times m\)
matrix whose rows form a basis of $\mathcal{A}_{x^\order}(\mat{F})$. We
implemented minor variants of the algorithms \algoname{M-Basis} (iterative, via
matrix multiplication) and \algoname{PM-Basis} (divide and conquer, via
polynomial matrix multiplication) from \cite{GiJeVi03}.  The lowest-level
function (\algoname{M-Basis-1} with the signature in \cref{algo:mbasis_one}),
handles order $\order=1$ in time $\bigO{\mathrm{rank}(\mat{F})^{\expmm-2}mn}$;
here, working modulo $X$, the matrix \(\mat{F}\) is over $\field$. Our
implementation follows \cite[Algo.\,1]{JeNeVi2018}, which returns an
\(\shifts\)-Popov basis, using only an additional row permutation compared to
the algorithm in \cite{GiJeVi03}.

\begin{algobox}
  \algoInfo{$\algoname{M-Basis-1}(\mat{F},\shifts)$}
  \label{algo:mbasis_one}

  \dataInfo{Input}{matrix $\mat{F}$ in $\matRing{m}{n}$, shift $\shifts$ in $\Z^m$}

  \dataInfo{Output}{the $\shifts$-Popov approximant basis for $(\mat{F},1)$}
\end{algobox}

This form of the output of \algoname{M-Basis-1} suffices to ensure that
\algoname{M-Basis} and \algoname{PM-Basis} return bases in \(\shifts\)-ordered
weak Popov form. Our implementation of \algoname{M-Basis} follows the original
design \cite{GiJeVi03}
with \(\order\) iterations, each computing the \emph{residual} \(\mat{R}\) and
updating \(\mat{P}\) via multiplication by a basis \(\mat{Q}\) obtained by
\algoname{M-Basis-1} on \(\mat{R}\). We also follow \cite{GiJeVi03} for
\algoname{PM-Basis}, using a threshold $T$ such that \algoname{M-Basis} is
called if $\order \le T$.  Building \algoname{PM-Basis} directly upon
\algoname{M-Basis-1}, i.e.  choosing \(T=1\), has the same cost bound but is
slower in practice.

\begin{algobox}
  \dataInfo{Input}{matrix $\mat{F}$ in $\pmatRing{m}{n}$, order \(\order\) in
  \(\ZZp\), shift $\shifts$ in $\Z^m$}

  \dataInfo{Output}{an $\shifts$-ordered weak Popov approximant basis for $(\mat{F},\order)$}

  \medskip

  \algoInfo{$\algoname{M-Basis}(\mat{F},\order,\shifts)$}
  \label{algo:mbasis}
  \algoSteps{
  \item \(\mat{P} \assign\) identity matrix in \(\pmatRing{m}{m}\), and \(\shiftss \assign \text{copy of } \shifts\)
  \item \algoword{For} $k = 0,\ldots,\order-1$:
    \begin{enumerate}[{\bf a.}]
      \item \(\mat{R} \in \matRing{m}{n} \; \assign\) coefficient of $\mat{P}\mat{F}$ of degree $k$
      \item \(\mat{Q} \in \pmatRing{m}{m} \; \assign \algoname{M-Basis-1}(\mat{R},\shiftss)\)
      \item \(\mat{P} \assign \mat{Q} \mat{P}\), and then \(\shiftss \assign \rdeg{\shifts}{\mat{P}}\)
    \end{enumerate}
  \item \algoword{Return} \(\mat{P}\)
  }
\end{algorithm}

\medskip

\algoInfo{$\algoname{PM-Basis}(\mat{F},\order,\shifts)$}
\label{algo:pmbasis}
\algoSteps{
\item \algoword{if} $\order \le T$ \algoword{return} $\algoname{M-Basis}(\mat{F},\order,\shifts)$
\item $\mat{P}_1 \assign \algoname {PM-Basis}(\mat{F} \bmod x^{\lceil \order/2\rceil},  \lceil \order/2\rceil, \shifts)$
\item $\mat{R} \assign \algoname{MiddleProduct}(\mat{P}_1, \mat{F}, \lceil \order/2\rceil, \lfloor \order/2\rfloor)$
\item $\shiftss \assign \rdeg{\shifts}{\mat{P}_1}$
\item $\mat{P}_2 \assign \algoname{PM-Basis}(\mat{R}, \lfloor \order/2\rfloor, \shiftss)$
\item \algoword{return} $\mat{P}_2 \mat{P}_1$
}
\end{algobox}
\noindent These algorithms use $\bigO{(m^\expmm + m^{\expmm-1}n)d^2}$ and $O((1+\frac{n}{m})
\MM'(m,d))$, respectively \cite{GiJeVi03}. Some implementation details are
discussed in \cref{ssec:intbas}. The next table compares timings for LinBox'
and our implementations of \algoname{PM-Basis} for a 20 bit FFT prime (LinBox'
implementation is not optimized for large primes and general primes).

{\small \hspace{0.6cm}
\begin{tabular}{c|c|c|ccc}
  $m$ & $n$ & $d$ & ours & Linbox & ratio \\
  \hline
  8	  & 4	  & 131072	& \textbf{6.6754}	& 15.5743	& 0.43 \\
  32	& 16	& 8192 &	\textbf{4.4185}	& 7.1150	& 0.62 \\
  128	& 64  & 2048 &	\textbf{18.0030}	& 28.7113	& 0.63 \\
  512	& 256 & 256	 &  \textbf{39.6255}	& 42.4051	& 0.93
\end{tabular}
}

\medskip

We also implemented \cite[Algo.\,3]{JeNeVi2018} which returns $\shifts$-Popov
bases and is about twice slower than \algoname{PM-Basis}; making this overhead
negligible for some usual cases is future work. For completeness, we handle
general approximants (with one modulus per column of $\mat{F}$) by an iterative
approach from \cite{BarBul92,BecLab94}; faster algorithms are more complex
\cite{JeNeScVi16,JeNeScVi17,JeNeVi2018} and use partial linearization
techniques. 

These techniques from \cite{Storjohann06,ZhoLab12} yield cost bounds in
\(\softO{m^{\expmm-1} n \order}\), which is a \(\Theta(\frac{m}{n})\) speedup
compared to \algoname{PM-Basis}. Implementing them is work in progress.
experimental code, which focuses for simplicity on \(n=1\) and ``generic''
inputs for which the degrees in \(\mat{P}\) can be predicted, revealed
significant speedups:

{\small\hspace{0.2cm}
\begin{tabular}{c|c|c|c|c}
  $m$ & $n$ & $\order$ & \algoname{PM-Basis} & \algoname{PM-Basis} with linearization \\
  \hline
  4   & 1 & 65536 & 1.6693 & \textbf{1.26891}\\
  16  & 1 & 16384 & 1.8535  & \textbf{0.89652} \\
  64	& 1 & 2048	& 2.2865	& \textbf{0.14362}\\
  256	& 1 & 1024	& 36.620	& \textbf{0.20660}
\end{tabular}
}

\medskip

Approximant bases are often applied to solve block-Hankel systems
\cite{LaChCa90}. In two specific settings, we have compared this approach to
the one which uses structured matrix algorithms; we are not aware of previous
comparisons of this kind. We obtain the following running times, using the
NTL-based solver from \cite{HyLeSc17}.

{\small\hspace{0.5cm}
\begin{tabular}{c|c|cc|cc}
  &     & \multicolumn{2}{c|}{Setting 1} & \multicolumn{2}{c}{Setting 2} \\
  $m$ & $d$  & \algoname{PM-Basis} & solver & \algoname{PM-Basis} & solver \\
  \hline
  5   & 8000 & \textbf{0.996}    & {8.23} & \textbf{2.19} & 3.820 \\
  12  & 1000 & \textbf{0.687}    & {6.18} & 2.33 & 2.28 \\
  30  & 500  & \textbf{2.84}     & {42.5} & 19.5 & \textbf{11.5}
\end{tabular}
}

\smallskip

Setting 1: we call \algoname{PM-Basis} on $\trsp{[\trsp{\mat{F}} \;\;
-\ident{m}]}$ at order $2d$ with shift \((0,\ldots,0)\), where $\mat{F}$ is an
$m\times m$ matrix of degree $2d-1$, and we solve a system with $m\times m$
Hankel blocks of size $d\times d$ (the structured solver returns a random
solution to the system). Our experiments show a clear advantage for approximant
algorithms. The asymptotic costs being similar, the effects at play here are
constant factor differences: approximant basis algorithms seem to be somewhat
simpler and to better leverage the main building blocks (matrix arithmetic over
\(\field\) and univariate polynomial arithmetic).

Setting 2 is the vector rational reconstruction problem. We call
\algoname{PM-Basis} on $\trsp{[\trsp{\mat{F}} \;\; -\ident{m}]}$ at order
$(m+1)d$ with shift \((0,\ldots,0)\), where $\mat{F}$ is a \(1\times m\) vector
of degree $(m+1)d-1$, and we solve a block system with $1\times m$ Hankel
blocks of size $md\times d$. The cost bounds are \(\softO{m^{\expmm+1} d}\) and
\(\softO{m^{\expmm}d}\), respectively. Approximants are faster up to dimension
about 15, which is explained by the arguments in the previous paragraph.  For
larger dimensions, as predicted by the cost estimates, the block-Hankel solver
is more efficient.


\paragraph{Interpolant bases.}
\label{ssec:intbas}

For matrices $\mat{E} = (\mat{E}_1,\dots,\mat{E}_\order)$ in $\K^{m\times n}$
and pairwise distinct points $\row{\alpha}=(\alpha_1,\dots,\alpha_\order)$ in
$\K$, consider
\[
  \mathcal{I}_{\row{\alpha}}(\mat{E}) =
  \{ \row{p}\in \K[x]^{1\times m} \mid
  \row{p}(\alpha_i)\mat{E}_i = 0 \text{ for } 1 \le i \le \order\}.
\]
An \emph{interpolant basis} for $(\mat{E},\row{\alpha})$ is a matrix whose rows
form a basis of the \(\polRing\)-module $\mathcal{I}_{\row{\alpha}}(\mat{E})$.
Note that
$\mathcal{I}_{\row{\alpha}}(\mat{F}(\alpha_1),\dots,\mat{F}({\alpha_\order}))$
coincides with $\mathcal{A}_{M}(\mat{F})$, for $\mat{F}$ in $\K[x]^{m \times
n}$ and $M=\Pi_{i=1}^{\order} (x-\alpha_i)$.

This definition is a specialization of those in \cite{BeLa2000, JeNeScVi17},
which consider \(n\) sets of points, one for each of the \(n\) columns of
$\mat{E}_1,\dots,\mat{E}_\order$: here, these sets are all equal. This more
restrictive problem allows us to give faster algorithms than those in these
references, by direct adaptations of the approximant basis algorithms presented
above. Besides, \cref{ssec:fracrec,sec:kernel} will show that interpolant bases
can often play the same role as approximant bases in applications.

These adaptations are described in \cref{algo:int_mbasis,algo:int_pmbasis},
where \(\row{\alpha}_{i\ldots j}\) stands for the sublist
\((\alpha_i,\alpha_{i+1},\ldots,\alpha_j)\).  In the next proposition, we
assume that $\MM(n,d)$ is in $\Omega(n^2 \M(d))$ (instead, one may add an extra
term $O(n^2 \M(d) \log(d))$ in the cost).

\begin{proposition}
  \cref{algo:int_pmbasis} is correct. For input evaluation points in geometric
  progression, it costs $O(\MM'(m,\order))$ if $n \le m$ and $O(\MM'(m,\order)+
  m^{\expmm-1} n \order \log(\order))$ otherwise.  For general evaluation
  points, an extra cost $O(m^2\M(\order) \log^2(\order))$ is incurred.
\end{proposition}
(Correctness follows from Items \emph{(i)} and \emph{(iii)} of
\cite[Lem.\,2.4]{JeNeVi2018}; the cost analysis is standard for such divide and
conquer algorithms.)

\begin{algobox}
  \dataInfo{Input}{
    matrices $\mat{E} = (\mat{E}_1,\dots,\mat{E}_\order)$ in $\K^{m\times n}$,
    evaluation points $\row{\alpha}=(\alpha_1,\dots,\alpha_\order)$ in \(\field\),
    shift $\shifts$ in $\Z^m$
  }

  \dataInfo{Output}{an $\shifts$-ordered weak Popov interpolant basis for $(\mat{E},\row{\alpha})$}

  \medskip

  \algoInfo{$\algoname{M-IntBasis}(\mat{E},\row{\alpha},\shifts)$}
  \label{algo:int_mbasis}

  \algoSteps{
  \item \(\mat{P} \assign\) identity matrix in \(\pmatRing{m}{m}\), and \(\shiftss \assign \text{copy of } \shifts\)
  \item \algoword{For} $k = 0,\ldots,\order-1$:
    \begin{enumerate}[{\bf a.}]
      \item \(\mat{R} \in \matRing{m}{n} \; \assign \mat{P}(\alpha_i) \mat{E}_i\)
      \item \(\mat{Q} \in \pmatRing{m}{m} \; \assign \algoname{M-Basis-1}(\mat{R},\shiftss)\)
      \item \(\mat{P} \assign \mat{Q} \mat{P}\), and then \(\shiftss \assign \rdeg{\shifts}{\mat{P}}\)
    \end{enumerate}
  \item \algoword{Return} \(\mat{P}\)
  }
\end{algorithm}

\medskip

\algoInfo{$\algoname{PM-IntBasis}(\mat{E},\row{\alpha},\tuple{s})$}
\label{algo:int_pmbasis}

\algoSteps{
\item \algoword{if} $\order \le T$ \algoword{return} $\algoname{M-IntBasis}(\mat{E},\row{\alpha},\shifts)$
\item $\mat{P}_1 \assign 
  \algoname {PM-IntBasis}(\mat{E}_{1\ldots\lceil \order/2 \rceil}, \row{\alpha}_{1\ldots\lceil \order/2\rceil},\shifts)$
\item \(\mat{R} \assign (\mat{P}_1(\alpha_{\lceil \order/2 \rceil +1})\mat{E}_{\lceil \order/2 \rceil +1},\ldots, \mat{P}_1(\alpha_\order) \mat{E}_\order)\)
\item $\shiftss \assign \rdeg{\shifts}{\mat{P}_1}$ 
\item $\mat{P}_2 \assign \algoname{PM-IntBasis}(\mat{R}, 
  \row{\alpha}_{\lceil \order/2 \rceil +1 \ldots \order}, \shiftss)$
\item return $\mat{P}_2 \mat{P}_1$
}
\end{algobox}

Our current code uses the threshold $T=32$ in the divide and conquer
\algoname{PM-Basis} and \algoname{PM-IntBasis}: beyond this point, they are
faster than the iterative \algoname{M-Basis} and \algoname{M-IntBasis}.  Unlike
in most other functions, where elements of \(\pmatRing{m}{n}\) are represented
as matrices of polynomials (\texttt{Mat<Vec<zz\_p>>} in NTL), in
\algoname{M-Basis} and \algoname{M-IntBasis} we see them as polynomials with
matrix coefficients (\texttt{Vec<Mat<zz\_p>>}). Indeed, since these algorithms
involve only matrix arithmetic over \(\field\) (recall that \(\deg(\mat{Q})\le
1\)), this turns out to be more cache-friendly and faster.

We implemented two variants for approximant bases: either the residual
\(\mat{R}\) is computed from \(\mat{P}\) and \(\mat{F}\) at each iteration, or
we initialize a list of residuals with a copy of \(\mat{F}\) and we update the
whole list at each iteration using \(\mat{Q}\). The second variant improves
over the first when \(n > m/2\), with significant savings when \(n\) is close
to \(m\). For interpolant bases, this does not lead to any gain.

Timings are showed in the next table, for Algorithms \algoname{M-Basis} (M),
\algoname{M-IntBasis} (M-I), \algoname{PM-Basis} (PM), \algoname{PM-IntBasis}
for general points (PM-I) and for geometric points (PM-Ig). For approximants,
we take a random input in $\K[x]^{m\times n}$ of degree $\order-1$; for
interpolants, we take $\order$ random matrices in $\K^{m\times n}$.
We focus on the common case $m \simeq 2n$, which arises for example in kernel
algorithms (\cref{sec:kernel}) and in fraction reconstruction, itself used in
basis reduction (\cref{sec:reduction}) and in the resultant algorithm of
\cite{Villard18} (\cref{sec:Villard}).

{\small\noindent\hspace{0.07cm}
  \setlength\tabcolsep{4.5pt}
\begin{tabular}{c|c||c|c|c||c|c|c|c}
  $m$ & $n$ & $\order$ & M & M-I & $\order$ & PM & PM-I & PM-Ig \\
  \hline
  4   &2  &32 & 1.60e-4 & 1.42e-4          & 32768 & \textbf{1.06}  & 6.81         & 1.47         \\
  16	&8	&32	& 1.98e-3 & \textbf{1.55e-3} & 4096  & \textbf{1.82}  & 5.51         & \textbf{1.92} \\
  32	&16	&32	&0.0104   & \textbf{7.59e-3} & 2048  & \textbf{3.90}  & 8.18         & \textbf{3.56} \\
  64	&32	&32	&0.0502   & \textbf{0.0354}  & 1024  & 8.1            & 12.2         & \textbf{6.38} \\
  128	&64	&32	&0.374    & \textbf{0.253}   & 1024  & 45             & 56.7         & \textbf{33.3} \\
  256 &128&32 &2.92     & \textbf{1.83}    & 1024  & 288            & 292          & \textbf{198}
\end{tabular}
}

\medskip

Concerning iterative algorithms, we observe that interpolants are slightly
faster than approximants, which is explained by the cost of computing the
residual \(\mat{R}\): it uses one Horner evaluation of \(\mat{P}\) and one
matrix product for interpolants, whereas for approximants it uses about
\(\min(k,\deg(\mat{P}))\) matrix products at iteration \(k\).

As for the divide and conquer algorithms, interpolant bases with general points are
slower, in some cases significantly, than the other two algorithms: although
the complexity analysis predicted a disadvantage, we believe that our
implementation of multipoint evaluation at general points could be improved to
reduce this gap. For the other two algorithms, the comparison is less clear.
There could be many factors at play here, but the main differences lie in the
base case (Step 1) which calls the iterative algorithm, and in the computation
of residuals (Step 3) which uses either middle products or geometric evaluation.
It seems that FFT-based polynomial multiplication performs slightly better than
geometric evaluation for small matrices and slightly worse for large matrices.


\vspace{-0.2cm}
\section{Higher-level algorithms}
\label{sec:higher}

In this section we consider kernel bases, system solving, determinants, and
basis reduction; we discuss algorithms which rely on multiplication, through
approximant/interpolant bases and lifting techniques. For many of these
algorithms, we are not aware of previous implementations or experimental
comparisons.


\paragraph{A note on matrix fraction reconstruction.}
\label{ssec:fracrec}

Given $\mat{H}$ in $\K(x)^{n \times n}$, a \emph{left fraction description} of
$\mat{H}$ is a pair of polynomial matrices $(\mat{Q},\mat{R})$ in $\K[x]^{n
\times n}$ such that $\mat{H} = \mat{Q}^{-1}\mat{R}$.  It is {\em minimal} if
$\mat{Q}$ and $\mat{R}$ have unimodular left matrix GCD and $\mat{Q}$ is in
reduced form (\emph{right fraction descriptions} are defined similarly).
Besides, $\mat{H}$ is said to be \emph{strictly proper} if the numerator of
each of its entries has degree less than the corresponding denominator.

Such a description of $\mat{H}$ is often computed from the power series
expansion of \(\mat{H}\) at sufficient precision, using an approximant basis.
Yet, for resultant computations in \cref{sec:resultant}, we would like to use
an interpolant basis to obtain this description from sufficiently many values
of \(\mat{H}\). We now state the validity of this approach; this is a matrix
version of rational function reconstruction \cite[Chap.\,5.7]{vzGathen13}. 

\begin{proposition}
  Let $\mat{H}$ be in $\K(x)^{n \times n}$ be strictly proper and suppose
  $\mat{H}$ admits left and right fraction descriptions of degrees at most $D$,
  for some $D\in\ZZp$. For $M$ in $\K[x]$ of degree at least $2D$ and such that
  all denominators in $\mat{H}$ are invertible modulo $M$, define the matrix
  $
    \mat{F} =
    \trsp{[ \mat{H} \bmod M \;\; -\ident{n} ]}
    \in \K[x]^{2n \times n}.
  $
  Then, if \(\mat{P} \in \pmatRing{2n}{2n}\) is a \(\tuple{0}\)-ordered weak
  Popov basis of $\mathcal{A}_{M}(\mat{F})$, the first \(n\) rows of
  \(\mat{P}\) form a matrix $[\mat{Q}\;\;\mat{R}]$ such that
  $(\mat{Q},\mat{R})$ is a minimal left fraction description of $\mat{H}$, with
  \(\mat{Q}\) in \(\tuple{0}\)-ordered weak Popov form.
\end{proposition}

The proof given in \cite[Lem\,3.7]{GiJeVi03} for the specific $M=x^{2D+1}$
extends to any modulus $M$; using an ordered weak Popov form (rather than a
reduced form) allows us both to know {\it a priori} that the first \(n\) rows are
those of degree at most \(D\), and to use degree \(2D\) instead of \(2D+1\)
(since \(\deg(\mat{R}) < \deg(\mat{Q})\) is ensured by this form).

In particular, if $M=\prod_{i=1}^{2D}(x-\alpha_i)$ for pairwise distinct points
\((\alpha_1,\ldots,\alpha_{2D})\), the interpolant basis algorithms in
\cref{ssec:intbas} give a minimal left fraction description of $\mat{H}$ from
$\mat{H}(\alpha_1),\dots,\mat{H}(\alpha_{2D})$.


\paragraph{Kernel basis.}
\label{sec:kernel}

We implemented two kernel basis algorithms: the first one, based on
\cref{lem:kernel_direct}, finds the kernel basis from a single approximant
basis at sufficiently large order; the second one, from \cite{ZhLaSt12}, uses
several approximant bases at small order and combines recursively obtained
kernel bases via multiplication. With a minor modification and no performance
impact, the latter returns an \(\shifts\)-ordered weak Popov basis.  In both
cases, we designed variants which rely on interpolant bases instead of
approximant bases. 

\begin{lemma}
  \label{lem:kernel_direct}
  Let \(\mat{F}\) be in \(\pmatRing{m}{n}\) of degree \(d\ge 0\), let
  \(\shifts\) be in \(\N^m\), and let \(\delta\) in \(\ZZp\) be an upper bound
  on the \(\shifts\)-degree of any \(\shifts\)-reduced left kernel basis of
  \(\mat{F}\); for example, \(\delta = n d + \max(\shifts) - \min(\shifts) +
  1\). Let \(M\) be in \(\polRing\) of degree at least \(\delta+d\), and
  \(\mat{P}\) in \(\pmatRing{m}{m}\) be an \(\shifts\)-reduced basis of
  $\mathcal{A}_{M}(\mat{F})$. Then, the submatrix of \(\mat{P}\) formed by its
  rows of \(\shifts\)-degree less than \(\delta\) is an \(\shifts\)-reduced
  left kernel basis for \(\mat{F}\).
\end{lemma}


\noindent For a proof, see
{\small\url{https://hal.archives-ouvertes.fr/hal-01995873v1/document}}.

In particular, one may find \(\mat{P}\) via \algoname{PM-IntBasis} at
\(d+\delta\) points or via \algoname{PM-Basis} at order \(d+\delta\); for
\(n\le m\), this costs \(\bigO{\MM'(m,d+\delta)}\). The approximant-based
direct approach is folklore \cite[Sec.\,2.3]{ZhLaSt12}, yet explicit statements
in the literature focus on shifts linked to the degrees in \(\mat{F}\), with
better bounds \(\delta\) (see
\cite[Lem.\,3.3]{ZhLaSt12},\cite[Lem.\,4.3]{NeiRosSol18}).

The algorithm of \cite{ZhLaSt12} is more efficient, at least when the entries
of \(\shifts\) are close to the corresponding row degrees of \(\mat{F}\); for a
uniform shift, it costs \(\softO{m^{\expmm} \lceil nd/m\rceil}\) operations.
We obtained significant practical improvements over the plain implementation of
\cite[Algo.\,1]{ZhLaSt12} thanks to the following observation: if \(n\le m/2\),
for a vast majority of input \(\mat{F}\), the approximant basis at Step 2 of
\cite[Algo.\,1]{ZhLaSt12}, computed at order more than \(2s\), contains the
sought kernel basis. Furthermore, this can be easily tested by checking
well-chosen degrees, and then the algorithm can exit early, avoiding the
further recursive calls. We took advantage of this via the following
modifications: we use order \(2s+1\) rather than \(3s\) (see
\cite[Rmk.\,3.5]{ZhLaSt12} for a discussion on this point), and when \(n>m/2\)
we directly reduce the number of columns via the divide and conquer scheme in
\cite[Thm.\,3.15]{ZhLaSt12}.

The use of approximants here follows the idea in \cref{lem:kernel_direct}: row
vectors of small degree which are in $\mathcal{A}_{M}(\mat{F})$ for a large
degree \(M\) must be in the kernel of \(\mat{F}\). Thus, one can directly
replace approximant bases with interpolant bases in \cite[Algo.\,1]{ZhLaSt12},
up to modifying Step 8 accordingly (dividing by the appropriate polynomial \(M\)).

Timings for both approaches are showed in the next table, for a random
\(\mat{F}\) of degree \(d\). Except for the last row, the shift is uniform and,
as expected, \cite[Algo.\,1]{ZhLaSt12} is faster than the direct approach; the
differences between interpolant and approximant variants follow those observed
in \cref{sec:appint}. The last row corresponds to a shift yielding the kernel
basis in Hermite form and shows, as expected, that the direct approach is
faster for shifts that are far from uniform.

We note that \cite[Algo.\,1]{ZhLaSt12} may use partial linearization if it
computes matrix products with unbalanced degrees or approximant bases with \(n
\ll m\). We have not yet implemented this part of the algorithm, which may lead
to slowdowns for some rare inputs.

{\small\hspace{0.6cm}
\begin{tabular}{c|c|c|c|c|c|c}
      &     &      & \multicolumn{2}{|c|}{direct} & \multicolumn{2}{|c}{\cite[Algo.\,1]{ZhLaSt12}} \\
  $m$ & $n$ & $d$  & approx. & int. & approx. & int. \\
  \hline
  8	&4	&8192	&7.22	&6.60	&\textbf{2.16}	& \textbf{2.49} \\
  8	&7	&8192	&14.1	&14.4	&\textbf{4.64}	&5.63 \\
  32	&16	&1024	&86.3	&63.1	&\textbf{3.75}	&\textbf{3.51} \\
  32	&31	&1024	&142	&118	&\textbf{8.27}	&\textbf{8.09} \\
  128	&64	&256	&2720	&1827	&16.8	&\textbf{11.8} \\
  128	&127	&256	& >1h & >1h &43.8	&\textbf{35.6}\\
  16	&1	&512	&\textbf{5.68}	&\textbf{5.31}	&11.5	&11.2
\end{tabular}
}


\paragraph{Linear system solving.}
\label{ssec:linsolve}

For systems $\mat{A} \col{\upsilon} = \col{b}$, with $\mat{A}$ in $\K[x]^{m
\times n}$, $\col{b}$ in $\K[x]^{m\times 1}$ and $\col{\upsilon}$ in $\K(x)^{n
\times 1}$, we implemented two families of algorithms. The first one uses
lifting techniques, assuming $\mat{A}$ is square, nonsingular, with
\(\mat{A}(0)\) invertible; the algorithm returns a pair $(\col{u},f)$ in
$\pmatRing{n}{1} \times \polRing$ such that $\mat{A} \col{u} = f \col{b}$ and
\(f\) has minimal degree. The second one uses a kernel basis and works for any
input \(\mat{A}\); under the assumptions above, it has a similar output.

\subsubsection*{Lifting techniques.}
Under the above assumptions, our lifting algorithm is standard: if $\mat{A}$
and $\col{b}$ have degree at most $d$, we first compute the truncated inverse
$\mat{S} = \mat{A}^{-1} \bmod x^{d+1}$ by matrix Newton
iteration~\cite{Schulz1933}. Then, we use Dixon's algorithm~\cite{Dixon82} to
compute $\col{\upsilon} \bmod x^{2nd} = \mat{A}^{-1} \col{b} \bmod x^{2nd}$; it
consists of roughly $2n$ steps, each involving a matrix-vector product using
either $\mat{A}$ or $\mat{S}$. Then, vector rational reconstruction is applied
to recover \((\col{u},f)\) from \(\col{\upsilon}\). The cost of this algorithm
is $O(\MM(n, d))$ for the truncated inverse of $\mat{A}$ and $O(n^3 \M(d))$ for
Dixon's algorithm; overall this is in $\softO{n^3 d}$.

To reduce the exponent in $n$, Storjohann introduced the \emph{high-order
lifting} algorithm \cite{Storjohann03}. The core of this algorithm is the
computation of $\Theta(\log(n))$ slices $\mat{S}_0,\mat{S}_1,\dots$ of the
power series expansion of $\mat{A}^{-1}$, where the coefficients of $\mat{S}_i$
are the coefficients of degree $(2^i-1)d-2^i+1,\dots,(2^i+1)d-2^i-1$ in
$\mat{A}^{-1}$. These matrices are computed recursively, each step involving 4
matrix products; the other steps of the algorithm, that use these $\mat{S}_i$
to compute $\col{\upsilon} \bmod x^{2nd}$, are cheaper, so the runtime is
$O(\MM(n, d)\log(n)) \subset \softO{n^\expmm d}$.

\subsubsection*{Using kernel bases.}

For this second approach, let $\mat{A}$ be any matrix in $\K[x]^{m \times n}$
and $\col{b}$ be in $\K[x]^{m\times 1}$.  The algorithm simply computes
\(\mat{K} \in \pmatRing{(n+1)}{k}\), a right kernel basis of the augmented
matrix \([\mat{A} \;|\; \col{b}] \in \pmatRing{m}{(n+1)}\). The matrix
\(\mat{K}\) generates, via \(\K(x)\)-linear combinations of its columns, all
solutions \(\col{\upsilon} \in \K(x)^{n\times 1}\) to $\mat{A}
\col{\upsilon} = \col{b}$.

In particular, if $\mat{K}$ is empty (i.e. \(k=0\), which requires \(m \ge
n\)), or if the last row of \(\mat{K}\) is zero, then the system has no
solution. Furthermore, if \(\mat{A}\) is square and nonsingular, \(\mat{K}\)
has a single column \(\trsp{[\trsp{\col{u}} \;|\; f]}\), where \(\col{u} \in
\pmatRing{n}{1}\) and \(f \in \polRing\), with \(f\) of minimal degree
(otherwise, \(\mat{K}\) would not be a basis).

In this context, the fastest known kernel algorithm is
\cite[Algo.\,1]{ZhLaSt12}. To exploit it best, we choose the input shift
\(\tuple{s} = (\tuple{d}, d)\), where \(d = \deg(\col{b})\) and $\tuple{d} \in
\N^n$ is the tuple of column degrees of \(\mat{A}\) (zero columns of
\(\mat{A}\) are discarded while computing \(\tuple{d}\)).

\subsubsection*{Implementation.}

We implemented the approaches described above: lifting with Dixon's algorithm,
high-order lifting, and via kernel. The table below shows timings for randomly
chosen $m\times m$ matrix \(\mat{A}\) and \(m\times 1\) vector \(\col{b}\),
both of degree $d$. In this case the lifting algorithms apply (with high
probability). On such inputs, Dixon's algorithm usually does best. High-order
lifting, although theoretically faster, is outperformed, mainly because it
performs $\Theta(\log(n))$ matrix products (we will however see that this
algorithm still plays an important role for basis reduction).  The kernel based
approach is moderately slower than Dixon's algorithm, but has the advantage of
working without any assumption on \(\mat{A}\).

{\small\hspace{0.8cm}
\begin{tabular}{c|c|c|c|c}
  $m$&	$d$& Dixon & high-order lifting & kernel\\
  \hline
  16&	1024&	\textbf{1.53}&	2.39& 2.07 \\
  32&	1024&	\textbf{4.94}&	13.8& 9.45 \\
  64&	1024&	\textbf{19.5}&	94.7& 46.5 \\
  128&	512 &	\textbf{55.2}&	266 & 108  \\
\end{tabular}
}
\smallskip


\paragraph{Determinant.}
\label{ssec:det}

We implemented four algorithms, taking as input a square \(m \times m\) matrix
\(\mat{A}\).

The most basic one uses expansion by minors, which turns out to be the fastest
option up to dimension about 6. 

The second one assumes that we have an element \(\alpha\) in \(\field\) of
order at least \(2md+1\), and uses evaluation/interpolation at the geometric
progression \(1, \alpha^2, \ldots, \alpha^{2md}\); this costs \(\bigO{m^3 \M(d)
+ m^{\expmm+1}d}\) operations in \(\field\). For dimensions between 7 and about
20, it is often the fastest variant, sometimes competing with the third.

The third one consists in solving a linear system with random right-hand side
of degree \(d\); this yields a solution \((\col{u},f)\) with \(f\) the sought
determinant up to a constant \cite{Pan1988}. For dimensions exceeding 20, this
is the fastest method (assuming that \(\mat{A}(0)\) is invertible).

The last one, from \cite{LaNeZh17}, is based on triangularization and runs in
\(\softO{m^\expmm d}\). Our implementation currently only supports the generic
case where the Hermite form of \(\mat{A}\) has diagonal
\((1,\ldots,1,\det(\mat{A}))\), or in other words, all so-called row bases
computed in that algorithm are the identity. This allows us to circumvent the
temporary lack of a row basis implementation, while still being able to observe
meaningful timings. Indeed, we believe that timings for a complete
implementation called on such ``generic'' input will be similar to the timings
presented here in many cases of interest, where one can easily detect whether
the algorithm has made a wrong prediction about the row basis being identity
(for example if the input matrix is reduced, which is the case if it is the
denominator of a minimal fraction description).
This recursive determinant algorithm calls expansion by minors as a base case
for small dimensions; for larger dimensions, it is generally slightly slower
than the third method.

The timings below are for a random \(m \times m\) matrix \(\mat{A}\) of degree
\(d\).

{\small\hspace{0.3cm}
\begin{tabular}{c|c|c|c|c|c}
  $m$ & $d$ & minors          & evaluation    & linsolve      & triangular      \\
  \hline                                                                       
  4	& 65536 & \textbf{0.7751} & 2.014         & 8.460         & \textbf{0.7826} \\
  16	&4096 & \(\infty\)      & \textbf{4.14} & 5.023         & 7.38            \\
  32	&4096 & \(\infty\)      & 34.5          & \textbf{25.4} & 41.6            \\
  64	&2048 & \(\infty\)      & 127           & \textbf{68.6} & 100             \\
  128	& 512 & \(\infty\)      & 244           & \textbf{96.6} & \textbf{99.0}    
\end{tabular}
}
\smallskip


\paragraph{Basis reduction.}
\label{sec:reduction}

Our implementation of the algorithm of \cite{GiJeVi03} takes as input a
nonsingular matrix \(\mat{A} \in \pmatRing{m}{m}\) of degree \(d\) such that
\(\mat{A}(0)\) is invertible, and returns a reduced form of \(\mat{A}\).  The
algorithm first computes a slice \(\mat{S}\) of $2d$ consecutive coefficients
of degree about \(md\) in the power series expansion of \(\mat{A}^{-1}\), then
uses \algoname{PM-Basis} to reconstruct a fraction description \(\mat{S}^{-1} =
\mat{R}^{-1} \mat{Q}\), and then returns \(\mat{R}\). A Las Vegas randomized
version is mentioned in \cite{GiJeVi03}, to remove the assumption on
\(\mat{A}(0)\): we will implement it for large enough \(\field\), but for
smaller fields this requires to work in an extension of \(\field\), which is
currently beyond the scope of our work.

In our experiments, to create the input \(\mat{A}\), we started from a random
\(m\times m\) matrix of degree \(d/3\) (which is reduced with high
probability), and we left-multiplied it by a lower unit triangular matrix and
then by an upper one, both chosen at random of degree \(d/3\).
The following table shows timings for both steps, with the first step either
based on Newton iteration or on high-order lifting; the displayed total time is
when using the faster of the two. We conclude that for reduction, as opposed to
the above observations for system solving, it is crucial to rely on high-order
lifting. Indeed, it improves over Newton iteration already for dimension 8, and
the gap becomes quite significant when the dimension grows.

{\small \hspace{0.3cm}
  \begin{tabular}{c|c|c|c|c|c}
    $m$ & $d$ & Newton & high-order & reconstruct & total\\
    \hline
    4   & 24574 & \textbf{1.251} & 1.688 & 8.772 & 10.02 \\
    8   & 6142  & 2.617 &  \textbf{2.244} & 8.851 & 11.09 \\
    16  & 1534  & 4.457 & \textbf{3.044} & 8.506 & 11.55 \\
    32  & 382   & 11.147 & \textbf{4.858} & 7.977 & 12.83 \\
    64  & 94    & 30.62 & \textbf{5.509} & 5.833 & 11.34 \\
  \end{tabular}
}


\section{Applications to bivariate resultants}\label{sec:Villard}

We conclude this paper with algorithms originating from Villard's
recent breakthrough on computing the determinant of structured
polynomial matrices~\cite{Villard18}. Fix a field $\K$ and consider
the two following questions: computing the resultant of two
polynomials $F,G$ in $\K[x,z]$ with respect to $z$, and computing the
characteristic polynomial of an element $A$ in $\K[z]/(P)$, for some $P$
in $\K[z]$.

The second problem is a particular case of the former, since the characteristic
polynomial of $A$ modulo $P$ is the resultant of $x-A(z)$ and $P(z)$ with
respect to $z$, up to a nonzero constant. Let $n$ be an upper bound
on the degree in $z$ of the polynomials we consider, and $d$ be a bound on
their degree in $x$ (so in the second problem, $d=1$). Villard proved that for
generic inputs, both problems can be solved in $\softO{n^{2-1/\expmm}d}
\subset \softO{n^{1.58}d}$ operations in $\K$.  For the first problem, the
best previous bound is $\softO{n^2 d}$, obtained either by
evaluation/interpolation techniques or Reischert's
algorithm~\cite{Reischert97}. For the second problem, the previous record was
$\softO{n^{\expmm_2/2}}$, where $\expmm_2$ is the exponent of matrix
multiplication in size $(s, s) \times (s, s^2)$, with $\expmm_2/2 \le
1.63$~\cite{LeUr18}. Note that these bounds apply to all inputs.

We show how the work we presented above allows us to put Villard's ideas to
practice, and outperform the previous state of the art for large input sizes.
This is however not straightforward: in both cases, this required modifications
of Villard's original designs (for the second case, using an algorithm
from~\cite{NeSaScVi19}).


\paragraph{Overview of the approach.}
\label{ssec:Valgo}

In \cite{Villard18}, Villard designed the following algorithm to find the determinant of a matrix
\(\mat{P}\) over \(\polRing\).

\begin{algobox}
  \algoInfo{$\algoname{Determinant}(\mat{P},m)$}
  \label{algo:det}

  \dataInfo{Input}{nonsingular $\mat{P}$ in $\pmatRing{\nu}{\nu}$; parameter $m
  \in \{1,\ldots,\nu\}$}

  \dataInfo{Output}{$\det(\mat{P})$}

  \algoSteps{
  \item \label{algo:det:stepi}
    compute $\bar{\mat{H}} =\mat{H} \bmod x^{2\lceil \nu/m\rceil d + 1}$,
    where $d$ is the degree of \(\mat{P}\) and $\mat{H}$ is the $m \times m$
    top-right quadrant of $\mat{P}^{-1} \in \K(x)^{\nu \times \nu}$
  \item \label{algo:det:stepii}
    from $\bar{\mat{H}}$, find a minimal left fraction description $(\mat{Q},\mat{R})$ of
    $\mat{H}$
  \item \label{algo:det:stepiii}
    return $\det(\mat{Q})$
  }
\end{algobox}

The parameter $m$ is chosen so as to minimize the theoretical cost. The
correctness of the algorithm follows from the next properties, which do not
hold for an arbitrary nonsingular $\mat{P}$: the matrix $\mat{H}$ is strictly
proper and admits a left fraction description $\mat{H}=\mat{Q}^{-1} \mat{R}$
such that $\det(\mat{P})=\det(\mat{Q})$, for $\mat{Q}$ and $\mat{R}$ in
\(\pmatRing{m}{m}\) of degree at most $\lceil\nu/m\rceil d$ (see
\cref{ssec:fracrec} for definitions). In~\cite{Villard18}, they are proved to
hold for generic instances of the problems discussed here.

Once sufficiently many terms of the expansion of $\mat{H}$ have been obtained
in \cref{algo:det:stepi}, the denominator $\mat{Q}$ is recovered by an
approximant basis algorithm and its determinant is computed by a general
algorithm in \cref{algo:det:stepii,algo:det:stepiii}, which cost $\softO{m^{\expmm}
(\nu d/m)}$.

While the algorithm applies to any nonsingular matrix \(\mat{P}\) satisfying
the properties above, in general it does not improve over previously known
methods (see \cref{ssec:det}). Indeed, the fastest known algorithm for
obtaining $\bar{\mat{H}}$ costs \(\softO{\nu^\expmm d}\) operations via
high-order lifting (see for example \cite[Thm.\,1]{GuSaStVa12}).


However, sometimes $\mat{P}$ has some structure which helps to speed up the
first step. Villard pointed out that when $\mat{P}$ is the Sylvester matrix of
two bivariate polynomials, then $\mat{P}^{-1}$ is a Toeplitz-like matrix which
can be described succinctly as $\mat{P}^{-1}=\mat{L}_1 \mat{U}_1 + \mat{L}_2
\mat{U}_2$; here, $\mat{L}_1$, $\mat{L}_2$ (resp.~$\mat{U}_1$, $\mat{U}_2$) are
lower (resp.~upper) triangular Toeplitz matrices with entries in $\K(x)$.
Hence, we start by computing the first columns $\row{c}_1$ of $\mat{L}_1$ and
$\row{c}_2$ of $\mat{L}_2$ as well as the first rows $\row{r}_1$ of $\mat{U}_1$
and $\row{r}_2$ of $\mat{U}_2$, all of them modulo $x^{2 \lceil\nu/m\rceil d
+1}$; then, \(\bar{\mat{H}}\) is directly obtained via the above formula for
$\mat{P}^{-1}$, using $\softO{m \nu d}$ operations. Computing these rows and
columns is done by solving systems with matrices $\mat{P}$ and $\trsp{\mat{P}}$
and very simple right-hand sides~\cite[Prop.~5.1]{Villard18}, with power series
coefficients, in time $\softO{\nu^2 d/m}$.  

Altogether, taking $m = \nu^{1/\expmm}$ minimizes the cost, yielding the
runtime $\softO{\nu^{2-1/\expmm} d}$. In the case of bivariate resultants
described above, the Sylvester matrix of $F$ and $G$ has size $\nu=2n$, hence
the cost bound $\softO{n^{2-1/\expmm} d}$.


\paragraph{Resultant of generic bivariate polynomials.}
\label{sec:resultant}

We implemented the algorithm described in the previous section to compute the
resultant of generic $F,G$ in $\K[x,z]$; first experiments showed that
obtaining $\row{c}_1$, $\row{c}_2$, $\row{r}_1$, $\row{r}_2$ was a bottleneck.
These vectors have power series entries and are solutions of linear systems
whose matrix is the Sylvester matrix of $F$ and $G$ or its transpose: they were
obtained via Hensel lifting techniques, following~\cite[Ch.\,15.4]{vzGathen13}.

To get better performance, we designed a minor variant of Villard's algorithm:
instead of computing the power series expansion of $\mat{H}$ modulo $x^\delta$,
where \(\delta = 2\lceil \nu/m\rceil d + 1\), we compute values of \(\mat{H}\)
at $\delta$ points. We choose these points in geometric progression and use the
interpolant basis algorithm of \cref{ssec:intbas} to recover $\mat{Q}$ and
$\mat{N}$, as detailed in~\cref{ssec:fracrec}. The value of $\mat{H}$ at
$x=\alpha$ is computed following the same approach as above, but over $\K$
instead of $\K[[x]]$. In particular, our implementation directly relies on
NTL's extended GCD algorithm over $\K=\F_p$ to compute the vectors $\row{c}_1$,
$\row{c}_2$, $\row{r}_1$, $\row{r}_2$.

The next table compares our implementation to the direct approach via
evaluation/interpolation; note that the latter approach, while straightforward
conceptually, is the state of the art.

{\small\hspace{0.4cm}
\begin{tabular}{c|c|c}
  $n = d$ & Direct & Algo.~\ref{algo:det} \\
  \hline
  100	&\textbf{1.75}	& 3.48\\
  200	&\textbf{17.4}	& 29.3\\
  300	&\textbf{72.3}	& 106\\
  400	&182 	& 182\\ 
\end{tabular}
~~
\begin{tabular}{c|c|c}
  $n = d$ & Direct & Algo.~\ref{algo:det} \\
  \hline
  600	&797	& \textbf{653}\\
  700	&1343	& \textbf{1081}\\
  800	&2121	& \textbf{1388}\\
  900	&3203	& \textbf{1760}\\
\end{tabular}
}
\smallskip

For these running times, input polynomials were chosen at random with partial
degree $n$ both in $x$ and in $z$; such polynomials have total degree $2n$, and
their resultant has degree $2n^2$. The largest examples have quite significant
sizes, but such degrees are not unheard-of in applications, as for instance in
the genus-2 point counting algorithms of~\cite{GaHa00,GaSc04,GaSc12,AbGaSp18}.
Overall, with \(n=d\), we observe a crossover point around $n=400$.  Besides,
in a close match with the analysis above, the parameter $m$ was set to $\lceil
n^{0.4} \rceil$ since this gave us the best runtimes. As an example, for $d =
300$, the cost of each individual steps were 65s for computing structured
inversions, and 40s for obtaining \(\mat{Q}\) and its determinant, which is a
good balance.


\paragraph{Characteristic polynomial.}

We consider the computation of the characteristic polynomial of an element $A$
in $\K[z]/(P)$, for some monic $P$ in $\K[z]$ of degree \(n\). The algorithm we
implemented, and which we sketch below, is from~\cite{NeSaScVi19} and assumes
that \(A\) and \(P\) are generic.

As explained previously, this problem is a particular case of a bivariate
resultant, but we rely on another point of view that allows for a better
asymptotic cost. Indeed, the characteristic polynomial of $A$ modulo $P$ is by
definition the characteristic polynomial of the matrix $\mat{M}$ of
multiplication by $A$ modulo $P$. In other words, it is the determinant of the
degree-1 matrix $\mat{P}=x \mat{I} - \mat{M} \in \pmatRing{n}{n}$.

The genericity assumption ensures that \(\mat{M}\) is invertible, hence the
power series expansion of $\mat{P}^{-1}$ is $\sum_{k \ge 0} -\mat{M}^{-k-1}
x^k$. Here, we use the top-left $m \times m$ quadrant $\mat{H}$ of $\mat{P}^{-1}$;
it has
entries $h_{i,j} \in \K[[x]]$, where
\[
h_{i,j,k} := \mathrm{coeff}(h_{i,j}, x^k) = \mathrm{coeff}(-z^j A^{-k-1} \bmod P, z^i).
\]
for $0 \le i,j < m$ and for all \(k\ge 0\).

A direct implementation of this idea does not improve on the runtime given in
\cref{ssec:Valgo}, since it computes $A^{-k-1} \bmod P$ for all $0\le k <
\delta = 2 \lceil n/m \rceil$ and therefore costs $\Omega(n^2/m)$. It turns out
that baby-steps giant-steps techniques allow one to compute $h_{i,j,k}$ for $0
\le i,j < m$ and $0 \le k < \delta$ in $\softO{\delta^{(\expmm-1)/2} n + m
n}$ operations in $\K$. Taking $m=\lceil n^{1/3} \rceil$ minimizes the
overall cost, resulting in the runtime $\softO{n^{(\expmm+2)/3}} \subset
O(n^{1.46})$.

The following table compares our implementation to NTL's built-in
characteristic polynomial algorithm, with random inputs $A$ and $P$. For such
inputs, NTL uses Shoup's algorithm for power projection~\cite{Shoup94}, which
runs in time $\softO{n^{(\expmm+1)/2}}$.
\smallskip

{\small\hspace{0.3cm}
  \begin{tabular}{c|c|c|c||c|c|c|c}
    n & m & NTL & new & n & m & NTL & new \\
    \hline
    5000  & 5 & \textbf{0.143} & 0.225 & 60000	&10	&8.45	&{\bf 8.34}\\
    20000	&8	&{\bf 1.43}  &1.62 & 80000	&10	&16.6	&{\bf 12.1} \\
    40000	&8	&4.69	&{\bf 4.42} & 100000	&10     &23.1	&{\bf 17.4}
  \end{tabular}
}

\subsubsection*{Acknowledgements.} Hyun was supported by a Mitacs Globalink
award; Neiger was supported by CNRS/INS2I's program for young researchers;
Schost was supported by an NSERC Discovery Grant.

\bibliographystyle{ACM-Reference-Format}

\end{document}